\documentclass[%
 reprint,
 amsmath,amssymb,
 aps,
]{revtex4-2}

\usepackage{graphicx}% Include figure files
\usepackage{dcolumn}% Align table columns on decimal point
\usepackage{bm}% bold math
\begin{document}

\preprint{APS/123-QED}

\title{Strain- and doping-tunable optical resonance in Kekulé-Y graphene}

\author{Yawar Mohammadi}%
 \email{y.mohammadi@cfu.ac.ir}
\affiliation{%
 Department of Physics Education, Farhangian University, P.O. Box 14665-889, Tehran, Iran
}%

\date{\today}% It is always \today, today,

\begin{abstract}

We investigate the optical response of Kekulé-Y graphene under uniaxial strain and carrier doping. Using a low-energy effective Hamiltonian, we show that strain reshapes the low-energy electronic structure of the Kekulé-Y phase and induces Van Hove singularities at energies well below those of pristine graphene. Within the Kubo formalism, we calculate the optical conductivity and identify multiple anisotropic interband features, with a pronounced resonance arising from strain-induced Van Hove singularities. The pronounced resonance is strongly anisotropic and robust against moderate thermal broadening and disorder, providing a clear optical signature of Kekulé-Y ordering. We further derive analytical expressions for the low-energy optical conductivity and the Drude weight, providing a detailed characterization of the strain- and doping-dependent optical response. Our results establish strain engineering as an effective route for controlling valley-dependent optical properties in Kekulé-Y graphene, originating from the Kekulé-induced coupling of the Dirac valleys, and suggest feasible optical probes for the experimental identification of the Kekulé-Y phase.

\end{abstract}

\maketitle

%\tableofcontents

\section{\label{sec:Introduction}Introduction}

Chiral-symmetry breaking in graphene can occur through Kekulé bond-order patterns, which couple Dirac fermions from the inequivalent $K$ and $K'$ valleys. Two main realizations are Y-type (Kekulé-Y), preserving gapless Dirac dispersions, and O-type (Kekulé-O), which mixes valleys and opens a spectral gap~\cite{Gutierrez1,Bao1,Gamayun1}. The electronic properties of graphene under Kekulé distortions have been extensively studied theoretically, covering transport, optical, and collective phenomena~\cite{Andrade1,Wang1,Guan1,Santacruz1,Tijerina1}. In the Kekulé-Y phase, Wu~\textit{et al.} predicted ballistic transport with valley-dependent conduction channels~\cite{Wu1,Andrade2}. Valley-selective tunneling, resonant conduction in nanoribbons, and junctions have also been demonstrated~\cite{Wang2,Iurov1,Garcia1,Li1}. Optical conductivity, including magneto-optical response, as well as integer quantum Hall signatures, have been investigated~\cite{Herrera1,Mohammadi1,SantacruzG1,Mohammadi2,Mohammadi3}, alongside studies of dynamical polarization and plasmon dispersion~\cite{Herrera2,Alimohammadi1}. By contrast, Kekulé-O studies focus on gap-related phenomena, such as Floquet-induced band reconstruction, flat-band formation in moiré geometries, and domain-wall soliton–induced confinement~\cite{Mojarro1,Zeng1,Scheer1,Garcia2}.
Valley-dependent transport effects in both phases further highlight the sensitivity of valley physics to the underlying modulation~\cite{Wang1}.

Mechanical strain is a powerful, noninvasive tool for engineering low-dimensional materials~\cite{Naumis1,Si1}. It allows continuous tuning of properties without chemical modification, reshaping the band structure, generating direction-dependent transport channels, modifying optical absorption~\cite{Pereira1,Ni1,Chhikara1}, and affecting phonon dispersion and electron–phonon coupling~\cite{Pellegrino1,Mohammadi4,Mohiuddin1}. In Kekulé-modulated graphene, strain offers an additional control over the electronic structure. While in the Kekulé-O phase strain competes with the modulation and reduces the band gap, in the Kekulé-Y phase it generates anisotropic gauge fields that valley-dependently distort the Dirac cones, leading to direction-dependent Fermi velocities and modified interband optical transitions~\cite{Andrade3}.

Nevertheless, the combined effects of strain, valley coupling, and finite carrier doping on the optical response of Kekulé-Y graphene remain largely unexplored. Understanding this interplay is essential for fundamental studies of symmetry-breaking phenomena and practical applications in valleytronics and tunable optoelectronic devices. Strain engineering enables precise, direction-dependent control of optical and transport responses without chemical modification, motivating a study of how uniaxial strain and finite doping reshape the low-energy electronic structure and optical conductivity.

In this study, we investigate the impact of uniaxial strain and finite doping on the electronic structure and optical conductivity of Kekulé-Y graphene. Using a low-energy effective Hamiltonian, we show that strain alters the bands and produces pronounced Van Hove singularities at lower energies than in pristine graphene. Through the Kubo formalism, we calculate the optical conductivity, revealing anisotropic intra- and interband features. A notable resonance originates from strain-induced Van Hove singularities, whose frequency and spectral weight depend on strain magnitude, direction, and carrier density. Analytical expressions for the low-energy optical conductivity and Drude weight provide insight into the strain- and doping-dependent response. These results highlight the resonance's robustness against moderate disorder and thermal effects, providing a clear optical signature of Kekulé-Y ordering. They further demonstrate that strain provides an effective route to tune valley-dependent optical properties arising from the Kekul\'e-induced valley coupling.

In the remainder of this paper, we introduce the low-energy Hamiltonian for Kekulé-Y graphene and discuss the resulting band structure and optical conductivity in Sec.~\ref{sec:Model}. Section~\ref{sec:Results} presents numerical results for strain- and doping-dependent optical conductivity and the underlying physical mechanisms. Finally, Sec.~\ref{sec:Conclusions} summarizes the main conclusions. Additional technical details are provided in appendices~\ref{AppendixA} and \ref{AppendixB}, outlining the construction of the low-energy Hamiltonian and the derivation of the optical conductivity, respectively.

\section{\label{sec:Model}Theoretical Model and Formalism}

\begin{figure}[b]
\includegraphics[width=7.5cm,angle=0]{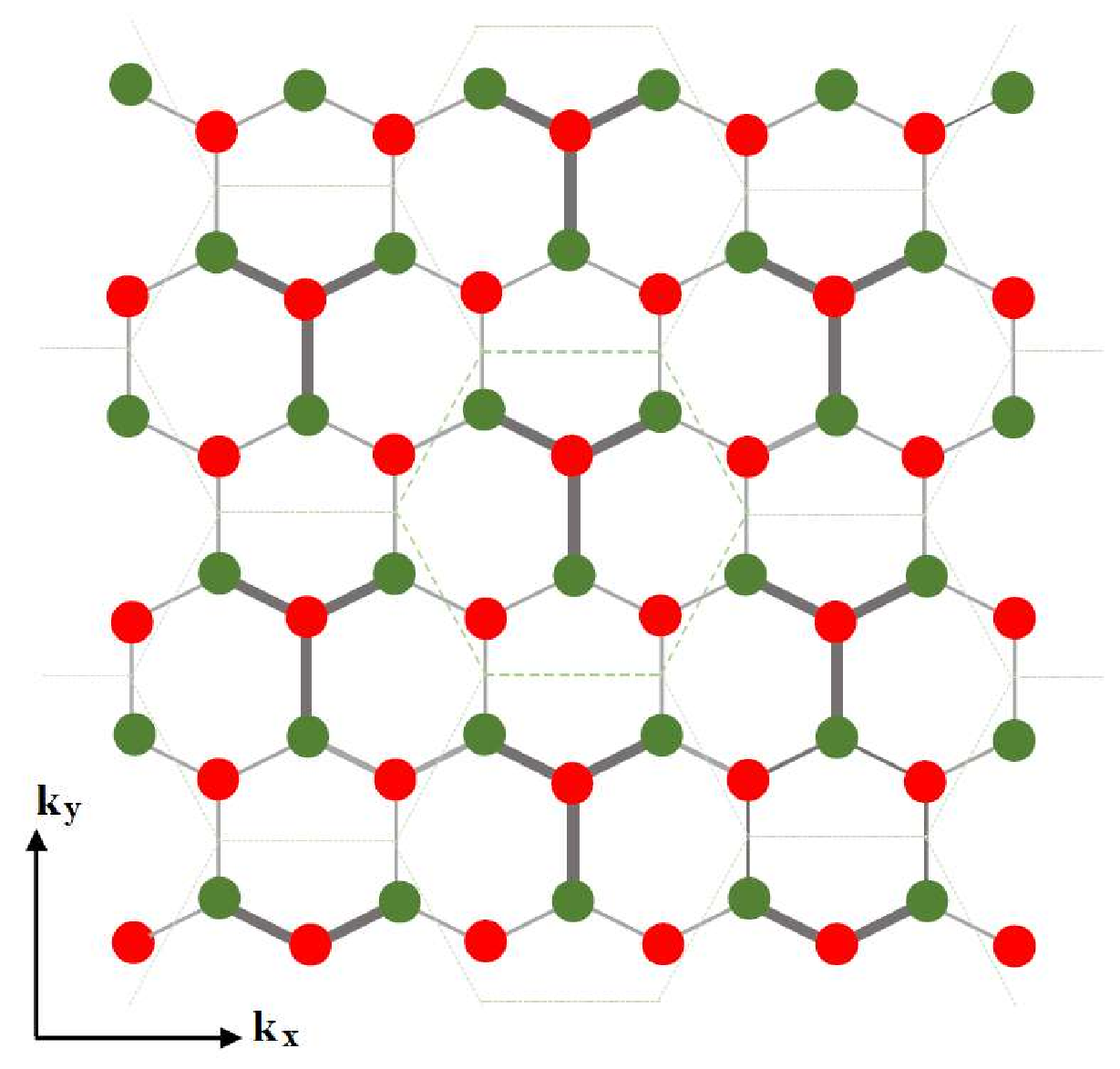}% Here is how to import EPS art
\caption{\label{Fig01} Schematic of a graphene lattice with a uniform Y-type Kekulé distortion. Dashed hexagons indicate the unit cell, and axes denote the reciprocal directions $k_x$ and $k_y$.}
\end{figure}

Fig.~\ref{Fig01} illustrates the unstrained lattice structure of Kekulé-Y graphene, highlighting the characteristic Kekulé-Y order formed by alternating strong and weak bonds. In this work, we study the electronic structure and optical response of Kekulé-Y graphene under a uniform uniaxial strain. To this end, we employ the continuum Hamiltonian derived in Ref.~\cite{Andrade3} and analyze the effects of finite doping and strain-induced anisotropy on the band structure and optical conductivity. In the following, we present the low-energy effective Hamiltonian, discuss the resulting band structure, and formulate the optical conductivity.

\subsection{\label{sec:Hamiltonian}Tight-binding Hamiltonian and energy spectrum}

The low-energy electronic properties of Kekulé-Y graphene under uniaxial strain can be described by~\cite{Andrade3}
\begin{equation}
\mathbf{H} = \left(
\begin{array}{cc}
\hbar v_{F}\bm{\sigma} \cdot (\bar{\bm{k}} + \bm{A}) & \mathbf{H}_{12} \\
\mathbf{H}_{12}^{\dag} & \hbar v_{F}\bm{\sigma} \cdot (\bar{\bm{k}} - \bm{A})
\end{array}
\right),
\label{Eq01}
\end{equation}
where the intervalley coupling is
\begin{equation}
\mathbf{H}_{12} = \Delta \left[ \hbar v_{F} (\bar{k}_x - i\bar{k}_y) \bm{\sigma}_0 + \hbar v_{F}(A_x - i A_y) \bm{\sigma}_z \right].
\label{Eq02}
\end{equation}
Here, \( \bm{\sigma}=(\bm{\sigma}_x,\bm{\sigma}_y) \), and \(v_F = \frac{3 a t_0}{2 \hbar}\) is the Fermi velocity, with nearest-neighbor hopping \(t_0 = 2.7~\textrm{eV}\) and carbon-carbon bond length \(a = 1.42~\textrm{\AA}\). The Pauli matrices \(\bm{\sigma}_i\) (\(i \in \{0,x,y,z\}\)) act on the sublattice pseudospin, and the Kekulé amplitude \(\Delta\) couples the two valleys.

Uniaxial strain enters the Hamiltonian through two distinct mechanisms. First, it deforms the crystal momentum according to\cite{Oliva1}
\begin{align}
\bar{\bm{k}} = \bigl[\bm{1} + (1-\beta)\bm{\epsilon}\bigr]\bm{k},
\label{Eq03}
\end{align}
where \(\bm{k}=(k_{x},k_{y})\) is the two-dimensional momentum in the unstrained lattice and \(\beta \approx 3\) is the Gr\"uneisen parameter characterizing the strain-induced correction to the hopping amplitudes~\cite{Pereira1}. The strain tensor, \( \bm{\epsilon}\), corresponding to a uniaxial deformation of magnitude \(\varepsilon\) which is applied at an angle \(\theta\) with respect to the zigzag direction (chosen as the \(x\)-axis), is given by
\begin{equation}
\bm{\epsilon} = \left(
\begin{array}{cc}
\varepsilon (\cos^2 \theta - \nu \sin^2 \theta) & \varepsilon (1 + \nu) \cos \theta \sin \theta \\
\varepsilon (1 + \nu) \cos \theta \sin \theta & \varepsilon (\sin^2 \theta - \nu \cos^2 \theta)
\end{array}
\right),
\label{Eq04}
\end{equation}
with Poisson ratio \(\nu \simeq 0.165\)~\cite{Blakslee1}. Second, strain induces a pseudovector potential \(\bm{A}\) that shifts the valley centers in momentum space and leads to a splitting of the Dirac cones. For the strain tensor \(\bm{\epsilon}\), this pseudovector potential takes the form
\begin{equation}
\mathbf{A} =
\left(
\frac{\beta}{2a}(\epsilon_{xx}-\epsilon_{yy}),
\; -\frac{\beta}{2a}\epsilon_{xy}
\right).
\label{Eq05}
\end{equation}
Eqs.~\ref{Eq01}-\ref{Eq05} fully specify the tight-binding description of strained Kekulé-Y graphene.

Diagonalizing Eq.~\ref{Eq01} yields
%\begin{widetext}
\begin{eqnarray}
E^{\lambda,s}_{\bm{k}} &=& \lambda\hbar v_{F} \frac{\sqrt{1+\Delta^{2}}}{\sqrt{2}} \Bigg[  |\bar{\bm{k}}-\bm{A}|^{2} + |\bar{\bm{k}}+\bm{A}|^{2}  + s \Bigg( \Big[ |\bar{\bm{k}}-\bm{A}|^{2} \nonumber \\
&-& |\bar{\bm{k}}+\bm{A}|^{2} \Big]^{2} + \Big[\frac{4 \Delta}{1+\Delta^{2}} |\bar{\bm{k}}-\bm{A}| |\bar{\bm{k}}+\bm{A}|\Big]^{2} \Bigg)^{1/2} \Bigg]^{1/2}.
\label{Eq06}
\end{eqnarray}
%\end{widetext}
Here, \(\lambda = \pm1\) labels conduction/valence bands, and \(s = \pm1\) distinguishes the strain-split branches.
In the unstrained limit (\(\varepsilon \to 0\)), where \(\bar{\bm{k}} = \bm{k}\) and \(\bm{A} = 0\), this reduces to the known Kekulé-Y dispersion~\cite{Gamayun1}.

\begin{figure}[b]
\includegraphics[width=8cm,angle=0]{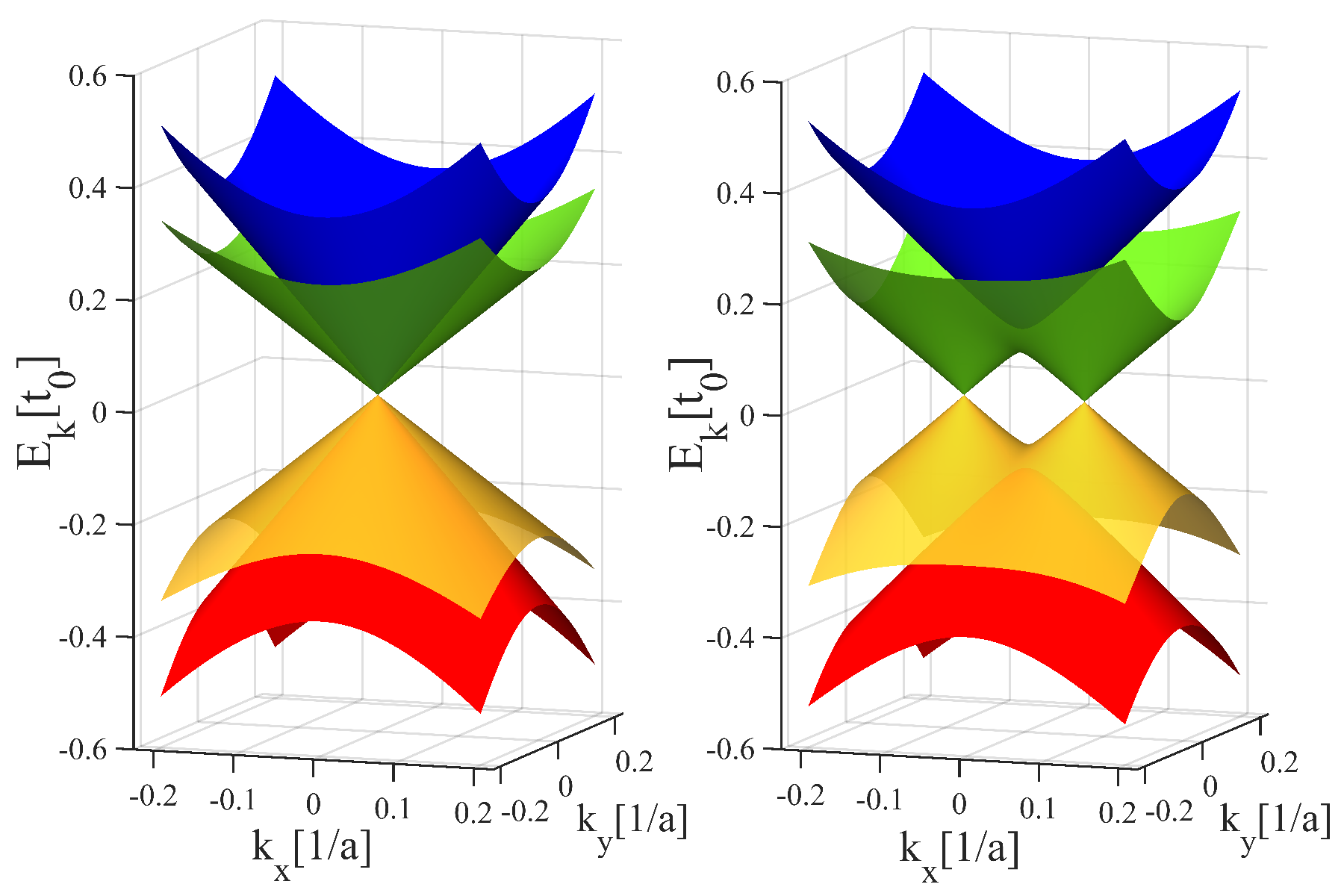}
\caption{\label{Fig02} Schematic 3D band structure of Kekul\'{e} graphene without strain (left) and under a uniaxial strain of 0.04 (right) applied along zigzag edge, illustrating the strain-induced modification of the electronic bands.}
\end{figure}

Fig. \ref{Fig02} shows the energy dispersion for both (a) the unstrained case and (b) the strained case, with $\Delta = 0.2$ and a uniaxial strain of magnitude $\varepsilon = 0.04$ applied along the $x$-axis (zigzag edge). As is evident from Fig.~\ref{Fig02} and Eq.~\ref{Eq06}, applying strain shifts the $E^{+,+}_{\bm{k}}$ and $E^{-,+}_{\bm{k}}$ bands upward and downward, opening a gap between them. Simultaneously, the $E^{+,-}_{\bm{k}}$ and $E^{-,-}_{\bm{k}}$ cones at the $\bm{\Gamma}$ point split into two at $\bar{\bm{k}} = \pm \bm{A}$, around which the band dispersion is linear in the strain-modified momentum $\bar{\bm{k}}$, $
E^{\lambda,-}_{\bm{k}} \approx \lambda\hbar v_F \frac{1-\Delta^{2}}{\sqrt{1+\Delta^{2}}}\, |\bar{\bm{k}} \mp \bm{A}| $~(see appendix~\ref{AppendixA}).

\begin{figure}[b]
\includegraphics[width=8cm,angle=0]{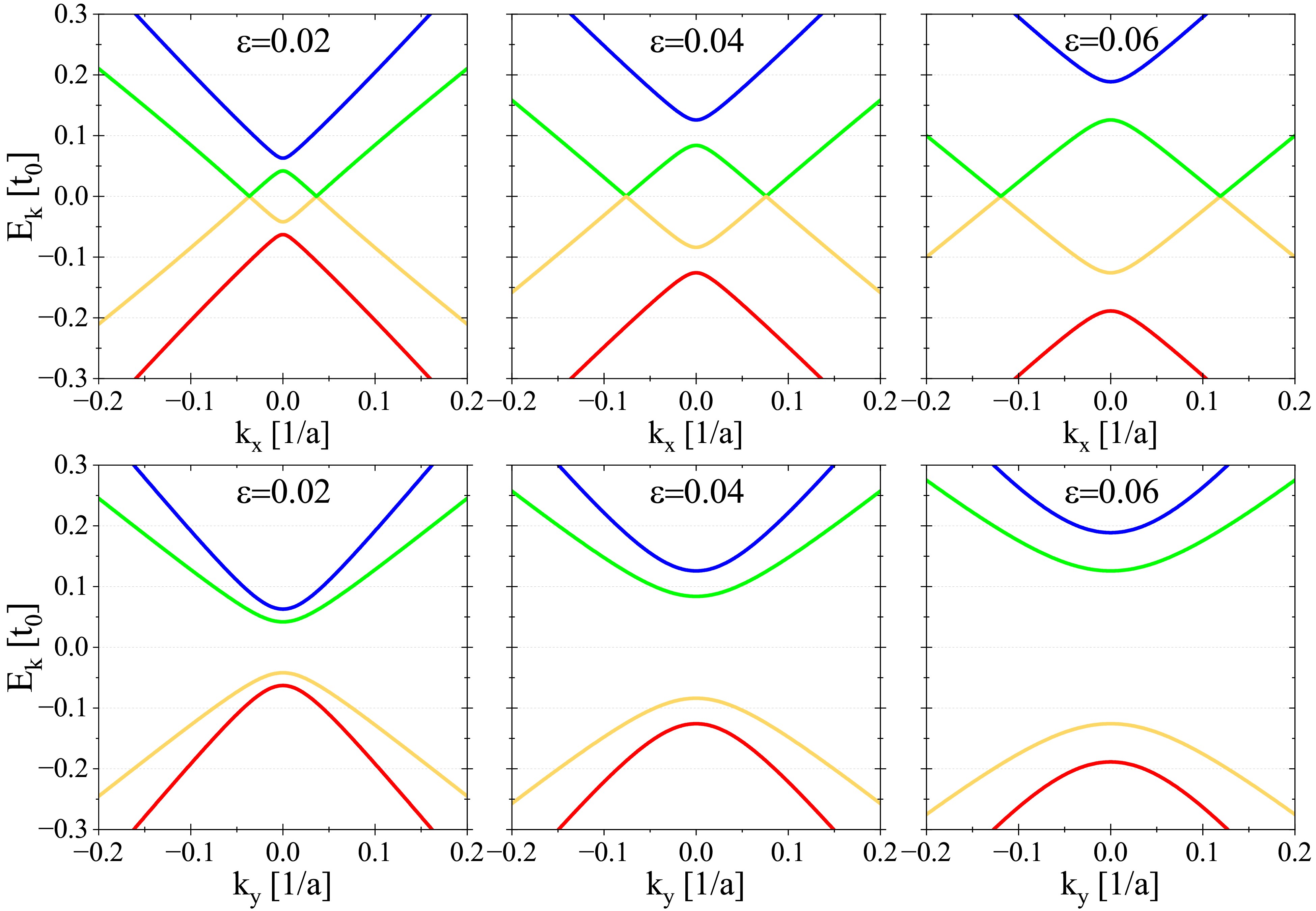}
\caption{\label{Fig03} Band-structure cuts along $k_y = 0$ (top) and $k_x = 0$ (bottom) for uniaxial strain values $\varepsilon = 0.02$, $0.04$, and $0.06$ (from left to right) applied along zigzag edge. Strain induces anisotropic dispersion and generates symmetric saddle points at $\boldsymbol{\Gamma}$, together with a strain-dependent gap to the next higher band.}
\end{figure}

To explore these strain-induced modifications in the band structure in more detail, we examine cross-sectional cuts of the band structure along $k_y = 0$ (upper panels) and $k_x = 0$ (lower panels), as presented in Fig.~\ref{Fig03}. The strain values considered are $\varepsilon = 0.02$, $0.04$, and $0.06$, which lie within the linear stress regime~\cite{Cao1}. Fig.~\ref{Fig03} and Eq.~\ref{Eq06} reveal that strain induces: (i) anisotropic dispersion, nearly linear along $k_x$ and parabolic along $k_y$; (ii) symmetric saddle points at the $\bm{\Gamma}$ point with energies $\pm \hbar v_F (1-\Delta) |\bm{A}|$, which give rise to Van Hove singularities in the density of states (DOS), thereby enhancing the joint density of states (JDOS) and influencing the optical and electronic response; and (iii) a gap opening to the next higher band, increasing as $2 \hbar v_F \Delta |\bm{A}|$.

\subsection{\label{sec:OpticalConductivity}Optical conductivity}

Within the framework of linear response theory, the optical conductivity is given by the retarded current-current correlation function, as described by the Kubo formula~\cite{Stauber1}
\begin{equation}
\sigma_{\alpha\beta}(\Omega+i0^{+})=
\frac{\Pi_{\alpha\beta}(\Omega+i0^{+})}{iA\hbar(\Omega+i0^{+})},
\label{Eq07}
\end{equation}
where $A$ is the sample area, $\Omega$ denotes the frequency, and  $\Pi_{\alpha\beta}(\Omega+i0^{+})$ represents the retarded current-current correlation function. The positive infinitesimal $0^{+}$ ensures causality and provides the correct analytic continuation, thereby allowing a clear separation of the real and imaginary parts. A brief overview of the formalism is presented here, while technical details are provided in appendix~\ref{AppendixB}.

The real part of the longitudinal optical conductivity can then be written as
\begin{equation}
\sigma_{\alpha\alpha}(\Omega)
=
D_{\alpha\alpha}\,\delta(\hbar\Omega)
+ \frac{\Im\Pi_{\alpha\alpha}(\Omega+i0^{+})}{A\hbar\Omega},
\label{Eq08}
\end{equation}
where $D_{\alpha\alpha}$ is the Drude weight (charge stiffness), defined by
\begin{equation}
D_{\alpha\alpha}
=
-\pi\lim_{\Omega\to0}
\frac{\Re\Pi_{\alpha\alpha}(\Omega+i0^{+})}{A\hbar}.
\label{Eq09}
\end{equation}
We note two important points. First, within the low-energy Dirac approximation, the diamagnetic contribution vanishes because the Hamiltonian is linear in momentum and contains no second-order momentum derivatives. Consequently, only the paramagnetic current–current correlation function contributes to the Drude weight, as commonly assumed for Dirac materials~\cite{Stauber2,Gusynin1,Peres1}. Second, the effects of short-range impurity scattering, leading to a finite spectral
broadening, will be discussed later..

The real and imaginary parts of the current--current correlation function are given by
\begin{eqnarray}\label{Eq10}
\Re\Pi_{\alpha\alpha}(\Omega + i0^{+}) &=&
g_{s}\hbar e^{2}\sum_{\mathbf{k}}\sum_{\lambda,\lambda',s,s'}  \\
&.& \frac{f_{\mathbf{k}}^{\lambda,s} - f_{\mathbf{k}}^{\lambda',s'}}
{\hbar\Omega + E_{\mathbf{k}}^{\lambda,s} - E_{\mathbf{k}}^{\lambda',s'}}
P_{\alpha\alpha}^{\lambda,s \to \lambda',s'}(\mathbf{k}), \nonumber
\end{eqnarray}
and
\begin{eqnarray}\label{Eq11}
\Im\Pi_{\alpha\alpha}(\Omega + i0^{+}) &=&
\pi g_s \hbar e^2 \sum_{\mathbf{k}} \sum_{\lambda,\lambda',s,s'}
\big[ f_{\mathbf{k}}^{\lambda,s} - f_{\mathbf{k}}^{\lambda',s'} \big] \\
&\delta& \left(\hbar \Omega + E_{\mathbf{k}}^{\lambda,s}
- E_{\mathbf{k}}^{\lambda',s'}\right)
P_{\alpha\alpha}^{\lambda,s \to \lambda',s'}(\mathbf{k}).  \nonumber
\end{eqnarray}

Here $f_{\mathbf{k}}^{\lambda,s}$ denotes the Fermi--Dirac distribution function, and
$P_{\alpha\alpha}^{\lambda,s \to \lambda',s'}(\mathbf{k})$ is the squared transition matrix element of the current operator, which characterizes the strength of electronic transitions between the initial $(\lambda,s)$ and final $(\lambda',s')$ states along the $\alpha$ direction. The derivation is provided in appendix~\ref{AppendixB}.

It should be noted that the optical conductivity calculated in this work corresponds to the total (valley-summed) response. In the Kekul\'e-Y phase, the original $K$ and $K'$ Dirac valleys are folded to $\Gamma$ point and coupled by the Kekul\'e modulation, so that the resulting eigenstates represent hybridized valley states. Consequently, the calculated optical conductivity describes the total response of these hybridized bands. Nevertheless, strain modifies the band structure in a valley-derived manner, which gives rise to anisotropic optical transitions reflected in the different components $\sigma_{xx}$ and $\sigma_{yy}$. In addition, the optical conductivity is evaluated in the long-wavelength limit ($q=0$), a standard approximation for optical measurements in graphene, since the photon momentum is negligible compared with the size of the Brillouin zone. Finite-$q$ effects may slightly modify optical transitions, especially for strain-shifted electronic states, but the main spectral features are expected to remain essentially unchanged. Studies including finite-$q$ corrections~\cite{Huang1} report only minor quantitative changes for small photon momenta, supporting the validity of the $q=0$ approximation.

In the following, we focus on interband processes and provide a qualitative discussion of the factors governing transition intensities in interband regime. We also briefly comment on the role of impurities. This overview offers physical insight into the mechanisms underlying the optical response and prepares the reader for the numerical evaluation of the optical conductivity presented in the next section.

\textit{Interband transitions at zero temperature}: At $T=0$, the Fermi-Dirac distributions reduce to step functions, and interband transitions are fully determined by two elements: the squared matrix element of the current operator and the energy-conserving Dirac delta function, \( \delta(\hbar \Omega + E_{\mathbf{k}}^{\lambda,s} - E_{\mathbf{k}}^{\lambda',s'}) \). When the \( P_{\alpha\alpha}^{\lambda,s \to \lambda',s'}(\mathbf{k}) \) varies smoothly across the Brillouin zone, its average value can be factored out of the \(\mathbf{k}\)-sum in Eq. \ref{Eq11} for each pair of valence and conduction bands \cite{Grosso1}. In this case, the frequency dependence of the optical response from that pair of bands is governed primarily by the joint density of states (JDOS),
\begin{equation}
\mathrm{JDOS}_{\lambda,s \to \lambda',s'}(\hbar \Omega) = \frac{g_s}{A} \sum_{\mathbf{k}} \delta(\hbar \Omega + E_{\mathbf{k}}^{\lambda,s} - E_{\mathbf{k}}^{\lambda',s'}),
\label{Eq12}
\end{equation}
which counts the number of available states satisfying the interband resonance condition. Critical points-maxima, minima, and saddle points, defined by \( \nabla_{\mathbf{k}}(E_{\mathbf{k}}^{\lambda,s} - E_{\mathbf{k}}^{\lambda',s'})=0 \), give rise to rapid variations of the JDOS.

Band-structure calculations show that uniaxial strain in strained Kekulé-Y graphene can generate band extrema and saddle points (see Fig.~\ref{Fig02}), thereby enhancing the density of states locally. As a consequence, the JDOS develops several pronounced peaks, including strongly enhanced Van Hove singularities at energies corresponding to these features. These singularities lead to intense interband optical transitions and produce prominent absorption peaks. Among them, those associated with the Van Hove singularities are particularly robust against moderate temperature variations and weak impurity scattering, as demonstrated in the next section.

The squared matrix elements \( P_{\alpha\alpha}^{\lambda,s \to \lambda',s'}(\mathbf{k}) \) govern the intensity of interband transitions and are influenced by the pseudospin structure of the bands. In systems with well-defined pseudospin chirality, the allowed or forbidden character of these transitions is determined by the chirality of the participating bands. For example, in pristine graphene, the pseudospin textures of conduction and valence bands permit interband transitions within a cone\cite{CastroNeto1}. In contrast, unstrained Kekulé-Y graphene exhibits a fully chiral fast cone and an anti-chiral slow cone. Therefore, current operator cannot induce a change in chirality, making interband transitions within a cone forbidden~\cite{Herrera1}. When strain is applied to Kekulé-Y graphene, critical points with locally gapped (massive) bands appear (see Fig.~\ref{Fig02}), where pseudospin chirality becomes ill-defined. Near these points, some interband transitions that are allowed in the unstrained bands are partially suppressed, while others that were forbidden may appear over a limited frequency range. This is similar to that reported in Kekulé-Y graphene due to on-site deviations~\cite{Mohammadi3}. In the present case, however, these effects primarily arise from strain. Moving away from the critical points, the bands gradually recover their approximate pre-strain chirality: the previously suppressed transitions strengthen, whereas the transitions that appeared near the gapped points weaken.

\textit{Thermal effects:}
To generalize the zero-temperature interband response to finite temperature, we replace the step-function occupations with the Fermi--Dirac distribution, \( f_{\mathbf{k}}^{\lambda,s} = [e^{\beta(E_{\mathbf{k}}^{\lambda,s}-\mu)} + 1]^{-1} \).
This accounts for thermal smearing of the occupation probabilities, particularly near the saddle points, softening the sharp JDOS peaks present at \(T=0\), while leaving the underlying band structure unchanged. Thermal effects are therefore included through the Fermi--Dirac occupations. Phonon-induced renormalization of the electronic spectrum is not explicitly considered in the present study; under the conditions considered, phonons are not expected to open a significant Dirac gap in Kekulé-Y graphene, unlike in Kekulé-O graphene where electron-phonon interactions can modify the gap~\cite{Li2}. Phonons may still slightly renormalize the band velocities and quasiparticle lifetimes, but no gap-opening occurs. Therefore, neglecting explicit phonon contributions is not expected to qualitatively modify the strain-induced optical features discussed in this work.

\textit{Impurity scattering:} Short-range disorder is incorporated through a phenomenological quasiparticle lifetime
$\Gamma$, leading to a finite spectral broadening of electronic transitions. In practice, the Dirac delta function in Eq.~(\ref{Eq08}) is replaced by a Lorentzian, \( \delta(x) \;\to\; \frac{\Gamma/\pi}{x^2 + \Gamma^2} \), which accounts for the broadening induced by point-like scatterers, such as vacancies or adsorbates~\cite{CastroNeto1}, while leaving the strain-modified band structure unchanged. This procedure is equivalent to the substitution $\Omega \rightarrow \Omega + i\Gamma/\hbar$ in the retarded current-current correlation function.

\textit{Intraband transitions}: The Drude weight (or charge stiffness), which characterizes the intraband optical response, is defined by Eqs.~\ref{Eq09} and \ref{Eq10} in the $\Omega \to 0$ limit. Therefore, only the intraband terms at the Fermi surface contribute to the Drude weight, since interband transitions require a finite excitation energy. Accordingly, the Drude weight takes the form
\begin{equation}
D_{\alpha\alpha}
= -\pi g_s \hbar e^2 \sum_{\mathbf{k}} \sum_{\lambda,s}
P_{\alpha\alpha}^{\lambda,s \to \lambda,s}(\mathbf{k}) \,
\frac{ \partial f_{\mathbf{k}}^{\lambda,s} }{ \partial E_{\mathbf{k}}^{\lambda,s} }.
\label{Eq13}
\end{equation}

For very small chemical potentials, where the Hamiltonian can be approximated by two Dirac-like Hamiltonians around the new Dirac points, \( \pm \mathbf{A} \), analytical expressions for the Drude weight can be obtained for both zero and finite temperatures (see appendix~\ref{AppendixA}).

At $T=0$ we have \( -\frac{ \partial f_{\mathbf{k}}^{\lambda,s} }{ \partial E_{\mathbf{k}}^{\lambda,s} } = \delta(E_{\mathbf{k}}^{\lambda,s}-\mu) \). Therefore, the Drude weight can be expressed analytically as
\begin{equation}
D_{\alpha\alpha} = \frac{g_s e^2}{2 \pi \hbar} \frac{v_\alpha}{v_{\bar{\alpha}}} |\mu|,
\label{Eq14}
\end{equation}
where $\bar{\alpha}$ is the direction perpendicular to $\alpha$ ($x \leftrightarrow y$). For strain applied along the zigzag direction, the Fermi velocities along the $x$ and $y$ directions are
\begin{subequations}
\begin{eqnarray}
v_x &= \hbar v_F \frac{1-\Delta^2}{\sqrt{1+\Delta^2}} \left[1 + (1-\beta)\varepsilon\right], \\
v_y &= \hbar v_F \frac{1-\Delta^2}{\sqrt{1+\Delta^2}} \left[1 - (1-\beta)\nu \varepsilon\right].
\end{eqnarray}
\label{Eq15}
\end{subequations}
These relations indicate that the Drude weight for an undoped strained Kekulé-Y lattice at $T=0$ is zero, and the anisotropy of the system is reflected in the different behaviors of the Fermi velocities along the $x$ and $y$ directions.

At finite temperature, thermal broadening of the Fermi distribution modifies the Drude weight, which is then given by
\begin{equation}
D_{\alpha\alpha} = \frac{g_s e^2}{2 \pi \hbar} \frac{v_\alpha}{v_{\bar{\alpha}}} \frac{\ln(1+e^{\beta \mu})}{\beta},
\label{Eq16}
\end{equation}
where $\beta = 1/k_B T$. In the limit $T \to 0$, we recover Eq.~\ref{Eq14} since \( \lim_{T \to 0} \frac{\ln(1+e^{\beta \mu})}{\beta} = |\mu| \).
At high temperature, $k_B T \gg \mu$, we have \( (k_B T)\ln(1+e^{\beta \mu}) \approx k_B T \ln 2 \), indicating that the Drude weight grows linearly with $T$, consistent with classical behavior. Our results for the strained Kekulé-Y lattice are consistent with previous studies of graphene under strain~\cite{Chhikara1,Pellegrino1}, and in the zero-strain limit they reduce to the well-known Drude weight of pristine graphene~\cite{Stauber1}.

From the analytical expressions in Eqs.~\ref{Eq14}-\ref{Eq16}, it is evident how strain modifies the Drude weight, and the quantitative impact for various parameters will be presented in the numerical results section.

\section{\label{sec:Results}Numerical Results and Discussion}

In this section, we present our results for the longitudinal conductivity, obtained through a numerical evaluation of Eqs.~\ref{Eq07}-\ref{Eq11}. The chemical potential $\mu$ is varied to capture a broad range of behaviors and to probe different regions of the band structure. For the numerical implementation of the terms containing Dirac delta functions for very weak impurity scattering, we use the Lorentzian representation $\delta(x) = \frac{\Gamma/\pi}{x^{2} + \Gamma^{2}}$, with a broadening parameter $\Gamma = 0.0005\, t_{0}$.

\begin{figure}[b]
\includegraphics[width=8cm,angle=0]{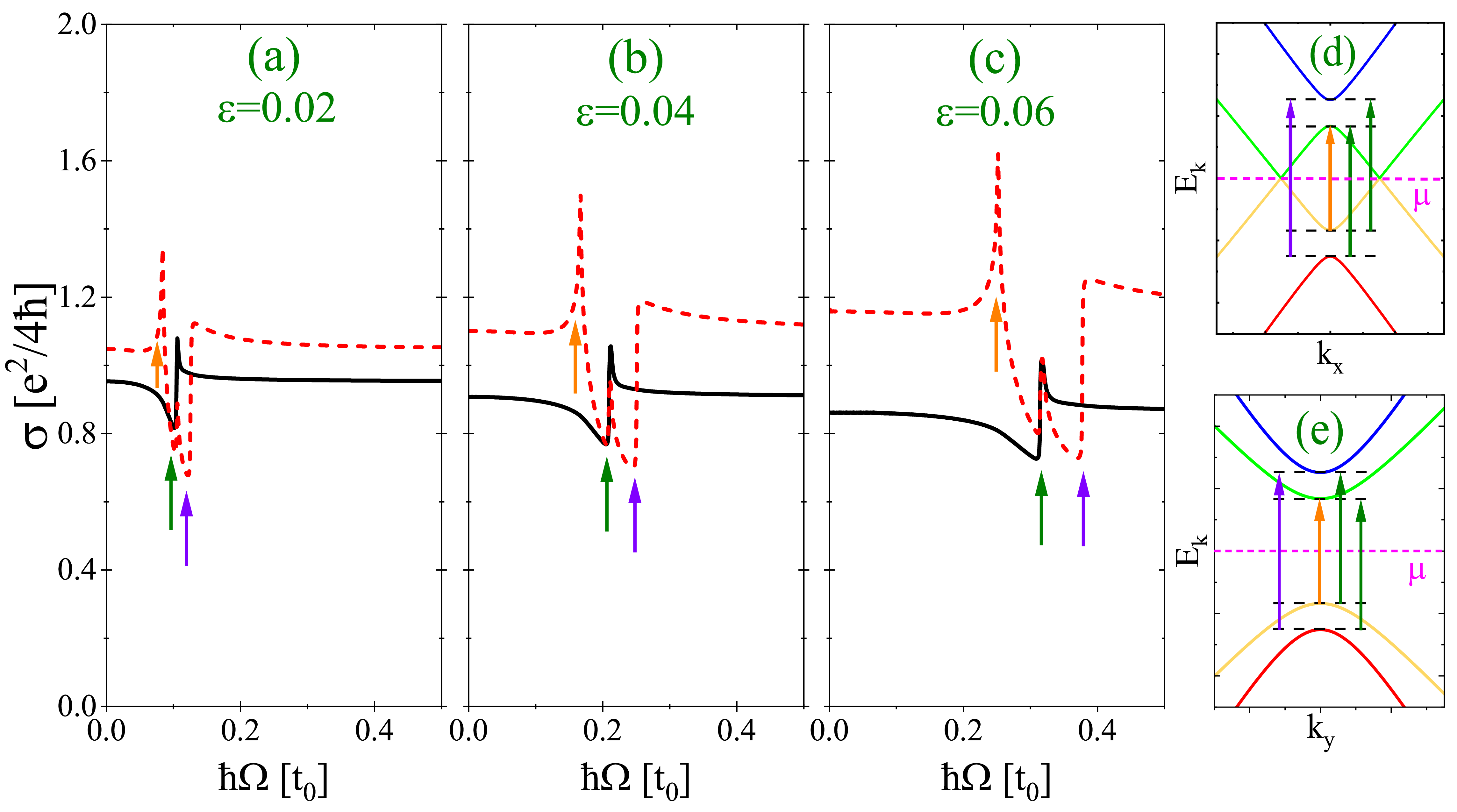}
\caption{\label{Fig04} Optical conductivity at $\mu = 0$ and $T = 0$ under uniaxial strain applied along the $x$ axis. Panels (a)-(c) present $\sigma_{xx}$ (black) and $\sigma_{yy}$ (red) for strain values of 0.02, 0.04, and 0.06, respectively. Panels (d) and (e) summarize the allowed absorptive interband transition groups; arrows on the band structures denote each transition group, while arrows of matching color mark their corresponding signatures in the conductivity.}
\end{figure}

Fig.~\ref{Fig04} illustrates the optical conductivity at $\mu=0$ and $T=0$ under uniaxial strain of $0.02$ [panel (a)], $0.04$ [panel (b)], and $0.06$ [panel (c)] along the $x$ axis. Black solid and red dashed curves indicate the conductivity along the $x$ and $y$ directions, respectively. Panels (d) and (e) summarize all allowed absorptive transition groups. Arrows on the band structures label each group, while arrows of matching color mark the corresponding features in the conductivity curves.

At low photon energies, the optical conductivity along the \( x \) and \( y \) directions remains nearly frequency-independent, reflecting the linear band dispersion around the new Dirac cones at \( \pm \mathbf{A} \). Although this flat response is analogous to that of pristine graphene, the magnitudes along the two directions differ due to the anisotropic Fermi velocities induced by strain. A detailed analysis based on the low-energy Hamiltonian, presented in appendix~\ref{AppendixA}, shows that the conductivities along the \( x \) and \( y\) directions for a uniaxial strain of magnitude $\epsilon$ applied along the $x$-axis (zigzag edge) are given by
\begin{subequations}
\begin{eqnarray}
\sigma_{xx} = \frac{v_x}{v_y} \frac{e^2}{4\hbar}
= \frac{1 + (1-\beta)\varepsilon}{1 - (1-\beta)\nu \varepsilon} \frac{e^2}{4\hbar},
\\
\sigma_{yy} = \frac{v_y}{v_x} \frac{e^2}{4\hbar}
= \frac{1 - (1-\beta)\nu \varepsilon}{1 + (1-\beta)\varepsilon} \frac{e^2}{4\hbar}.
\end{eqnarray}
\label{Eq17}
\end{subequations}
For a representative strain of \( \varepsilon = 0.04 \), these analytical expressions predict optical conductivities of
\( 0.91\, (e^2/4\hbar) \) and \( 1.10\, (e^2/4\hbar) \) along the \( x \) and \( y \) directions, respectively, in excellent agreement with the numerical results shown in the figure.

As the photon energy increases, \( \sigma_{xx} \) develops a pronounced dip–peak–like spectral feature, whereas \( \sigma_{yy} \) exhibits two consecutive peak–dip structures of larger intensity. These spectral structures appear in all three panels, with their positions shifting to higher energies as the applied strain increases from left to right. The locations and relative amplitudes of these features provide clear fingerprints of both the strain magnitude and the measurement direction. At higher photon energies, the conductivity evolves into a nearly flat, anisotropic regime, reflecting the reconstructed, unstrained band structure at high energies.

\begin{figure}[b]
\includegraphics[width=8cm,angle=0]{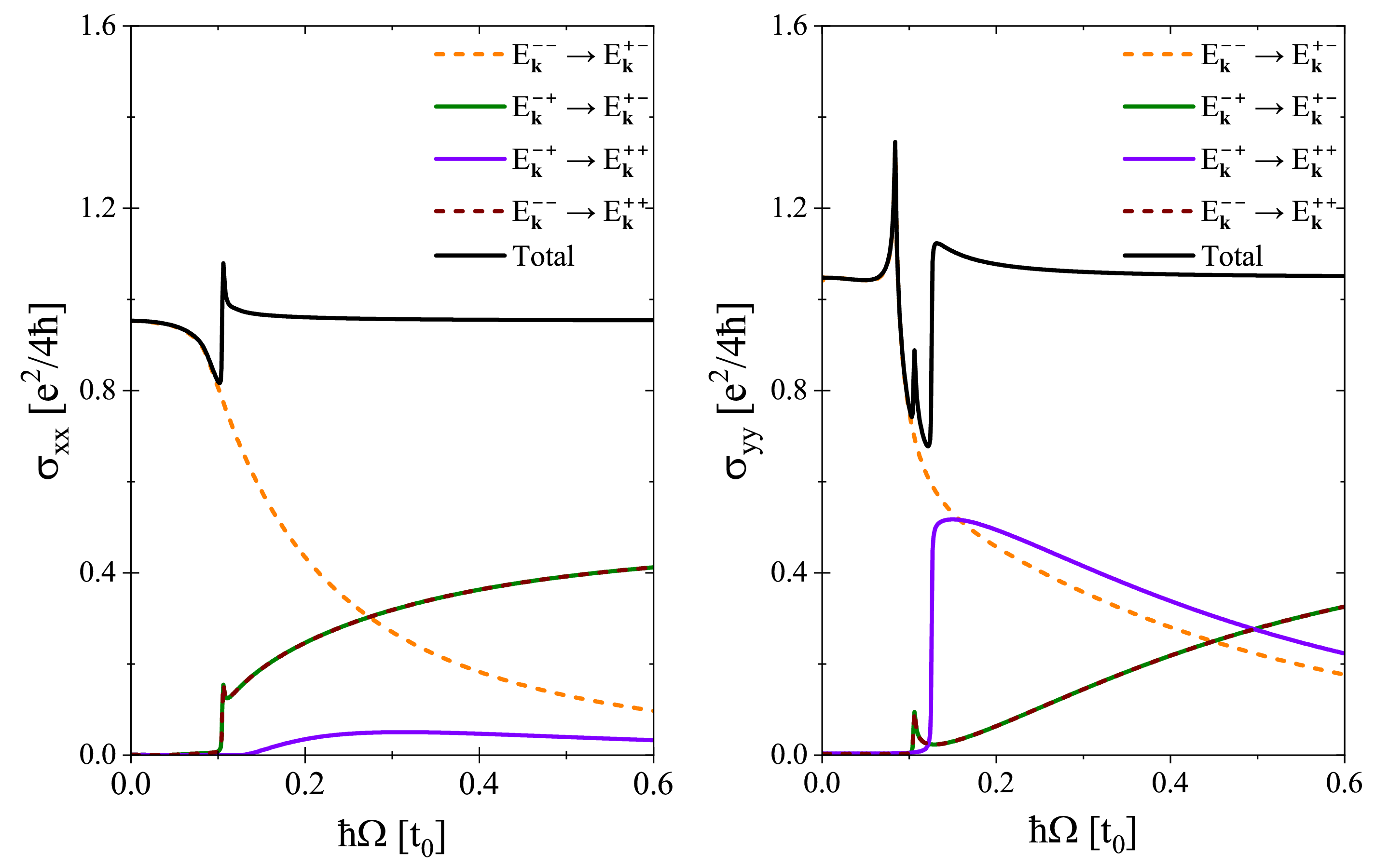}
\caption{\label{Fig05} Contributions of all allowed transition groups to the optical conductivity at $\mu=0$, $T=0$, under 0.02 uniaxial strain. Left and right panels correspond to $\sigma_{xx}$ and $\sigma_{yy}$, respectively; colored curves indicate individual transitions, and black curves represent the total conductivity.}
\end{figure}

To elucidate the origin of the optical conductivity features, Fig.~\ref{Fig05} decomposes the total conductivity into contributions from all possible interband transitions for a fixed strain of 0.04. The left panel displays $\sigma_{xx}$ together with all transition contributions, while the right panel shows $\sigma_{yy}$. Each transition curve matches the arrow marking the corresponding transition group in Fig.~\ref{Fig04}.

This figure confirms that the frequency-independent low-energy conductivity originates solely from the linear band around the Dirac cones at $\pm A$, i.e., from transitions between the $E^{--}_{\mathbf{k}}$ and $E^{+-}_{\mathbf{k}}$ bands. As the photon energy increases, these transitions lead to distinct behaviors along the $x$ and $y$ directions: $\sigma_{xx}$ decreases, whereas $\sigma_{yy}$ develops a pronounced peak. This anisotropic response arises from local band curvature near the Van Hove singularity, which enhances the JDOS along specific directions.

Upon further increasing the photon energy, additional interband transitions contribute to the optical conductivity. The sharp increase observed at the onset of some transitions originates from strain-induced extrema in the band structure. Band flattening or gap opening leads to a significant increase in the JDOS. At the same time, the intensity of each contribution is modulated by the corresponding squared velocity matrix elements, $P^{\lambda,s \to \lambda',s'}_{\alpha\alpha}(\mathbf{k})$, whose behavior and chirality dependence were discussed in the previous section~\ref{sec:OpticalConductivity}. For transitions that are chirality-allowed in unstrained Kekulé-Y graphene, strain enhances both the JDOS and the associated matrix elements, giving rise to pronounced initial peaks and a conductivity that continues to increase with photon energy. In contrast, for transitions that are chirality-forbidden in unstrained Kekulé-Y graphene, strain only partially relaxes the selection rules through band mixing~\cite{Chung1,Pereira2}. Consequently, although onset features may appear due to enhanced JDOS near extrema, their intensity remains comparatively weak.

At higher photon energies, the influence of strain on the band structure gradually diminishes, and the bands progressively recover their original chirality characteristics of the unstrained system\cite{Herrera1}. In this regime, contributions from transitions that are forbidden in unstrained Kekulé-Y graphene tend to vanish, while those from allowed transitions asymptotically approach their unstrained values. The competing increase and decrease of different interband contributions at high energies lead to an optical conductivity that becomes nearly constant and weakly frequency-dependent~\cite{Herrera2}.

\begin{figure}[b]
\includegraphics[width=8cm,angle=0]{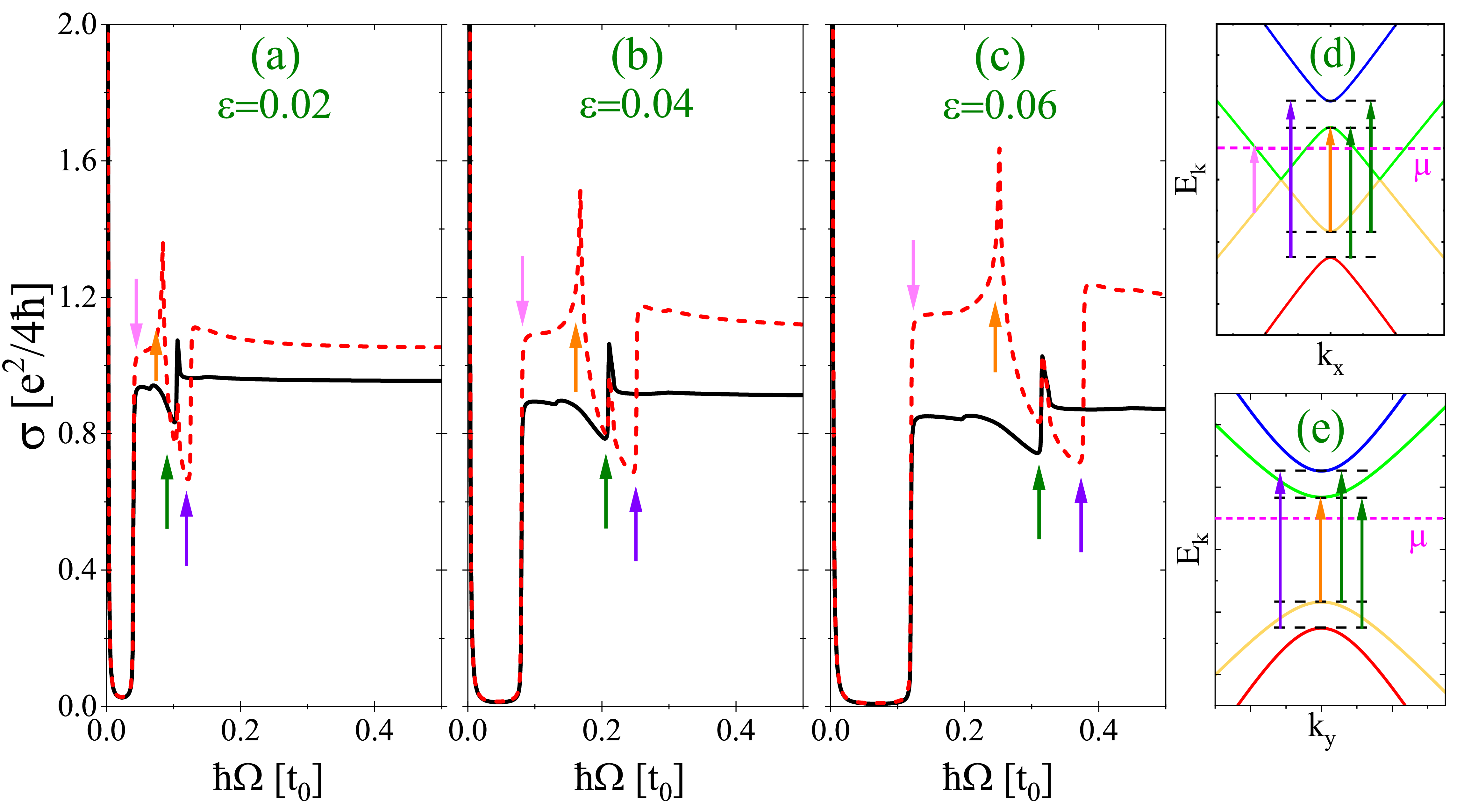}
\caption{\label{Fig06} Optical conductivity of strained Kekulé-Y graphene with the chemical potential set at half the saddle-point energy. Panels (a)–(c) show $\sigma_{xx}$ (black solid curves) and $\sigma_{yy}$ (red dashed curves) for three representative chemical potentials, $\mu = 0.02\,t_0$, $0.04\,t_0$, and $0.06\,t_0$, respectively. Panels (d) and (e) summarize the allowed absorptive interband transition groups; arrows on the band structures denote each transition group, while arrows of matching color mark their corresponding signatures in the conductivity.}
\end{figure}

Fig.~\ref{Fig06} presents the optical conductivity under the same as in Fig.~\ref{Fig04}, but with the chemical potential set at half the saddle-point energy. Panels~(a)–(c) display $\sigma_{xx}$ and $\sigma_{yy}$, indicated by black solid and red dashed curves, respectively, for $\mu=0.02~t_0$, $0.04~t_0$, and $0.06~t_0$. Arrows indicate allowed absorptive transitions as before.
At finite doping, a finite intraband (Drude) contribution emerges. Additionally, Pauli blocking prohibits interband absorption for photon energies $\hbar\Omega < 2\mu$ producing a low-frequency window of vanishing conductivity. This threshold originates from the cone-like band crossings~\cite{Stauber1} near the strain-induced Dirac points.

\begin{figure}[b]
\includegraphics[width=8cm,angle=0]{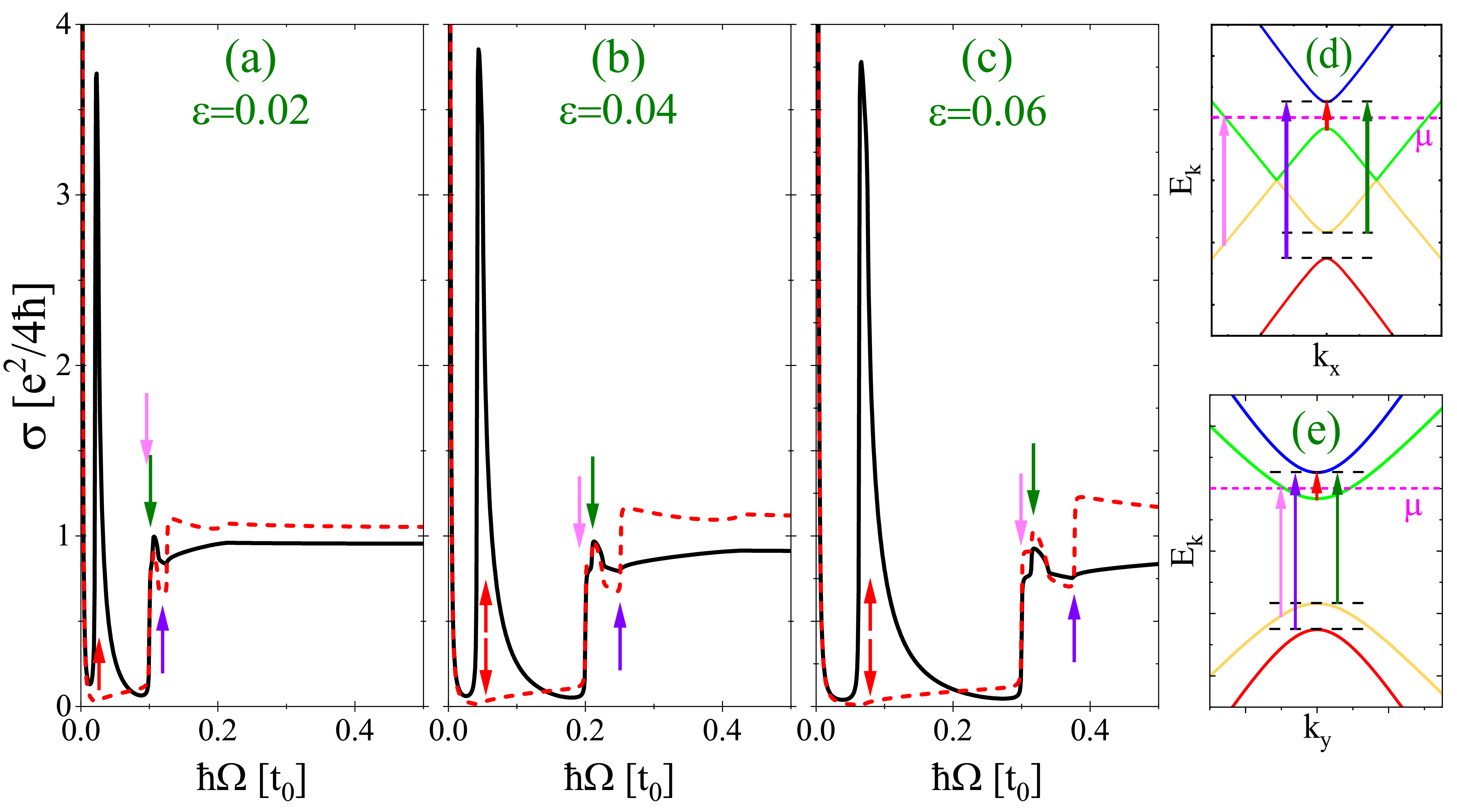}
\caption{\label{Fig07} Optical conductivity of strained Kekulé-Y graphene with the chemical potential set at the midpoint of the band gap above the saddle-point energy. Panels (a)–(c) show $\sigma_{xx}$ (black solid curves) and $\sigma_{yy}$ (red dashed curves) for three representative chemical potentials, $\mu = 0.05\,t_0$, $0.10\,t_0$, and $0.15\,t_0$, respectively. Panels (d) and (e) summarize the allowed absorptive interband transition groups; arrows on the band structures denote each transition group, while arrows of matching color mark their corresponding signatures in the conductivity.}
\end{figure}

Fig.~\ref{Fig07} shows the optical conductivity with the chemical potential set at the midpoint of the band gap above the saddle-point energy. Panels~(a)–(c) display $\sigma_{xx}$ and $\sigma_{yy}$, indicated by black solid and red dashed curves, respectively, for $\mu=0.05~t_0$, $0.10~t_0$, and $0.15~t_0$. Arrows indicate allowed transitions as before.

As is evident, three key modifications are observed: First, a finite intraband (Drude-like) contribution arises. Although the chemical potential lies within the local band gap above the saddle point, it intersects a continuation of the same band at other wave vectors, producing low-frequency spectral weight. Second, the distinct spectral features are partially suppressed by Pauli blocking transitions from $E^{-+}_{\mathbf{k}}$ to $E^{+-}_{\mathbf{k}}$, reducing the intensity of these features by approximately half. Consequently, the transitions limited by the $2\mu$ Pauli-blocking threshold produce a step-like onset in the optical conductivity. Third, a sharp, anisotropic optical resonance emerges in $\sigma_{xx}$, with much weaker and broader response along $\sigma_{yy}$. Its position shifts to higher energies with increasing strain, while its width slightly increases due to the energy range of transitions. These features are linked to transitions near the Van Hove singularity.

In realistic graphene-based systems, disorder arising from different types of impurities can influence the optical response through carrier scattering. Charged impurities associated with substrate ions or chemical residues generate long-range Coulomb potentials and mainly lead to small-angle scattering, whereas short-range defects such as vacancies, lattice imperfections, or adsorbed atoms can induce stronger momentum relaxation and intervalley scattering. Resonant impurities may also introduce localized electronic states near the Dirac point and modify optical transitions. The influence of such scattering mechanisms on the electronic transport and optical properties of graphene has been widely discussed in the literature~\cite{Peres2,Adam1,DasSarma1,Yuan1}.

These scattering processes limit the carrier lifetime $\tau$ and lead to a finite broadening of optical resonances, which is commonly described phenomenologically by a Lorentzian damping parameter $\Gamma\approx\hbar/\tau$ in optical response calculations~\cite{Peres2,DasSarma1}. Moreover, the spatial correlation of disorder can significantly affect carrier dynamics. When the impurity potential varies smoothly over a finite correlation length, intervalley scattering is typically suppressed and the carrier lifetime may increase compared with completely random disorder. As a consequence, correlated disorder may lead to narrower optical resonances and more pronounced spectral features~\cite{Li3}. As discussed in Ref.~\cite{Li3}, correlated disorder can influence carrier lifetimes and the sharpness of optical resonances, providing a qualitative example consistent with the general discussion above.

This phenomenological description sets the stage for examining disorder effects in our numerical simulations, where the broadening parameter $\Gamma$ is varied to mimic different impurity scattering strengths.

\begin{figure}[b]
\includegraphics[width=8cm,angle=0]{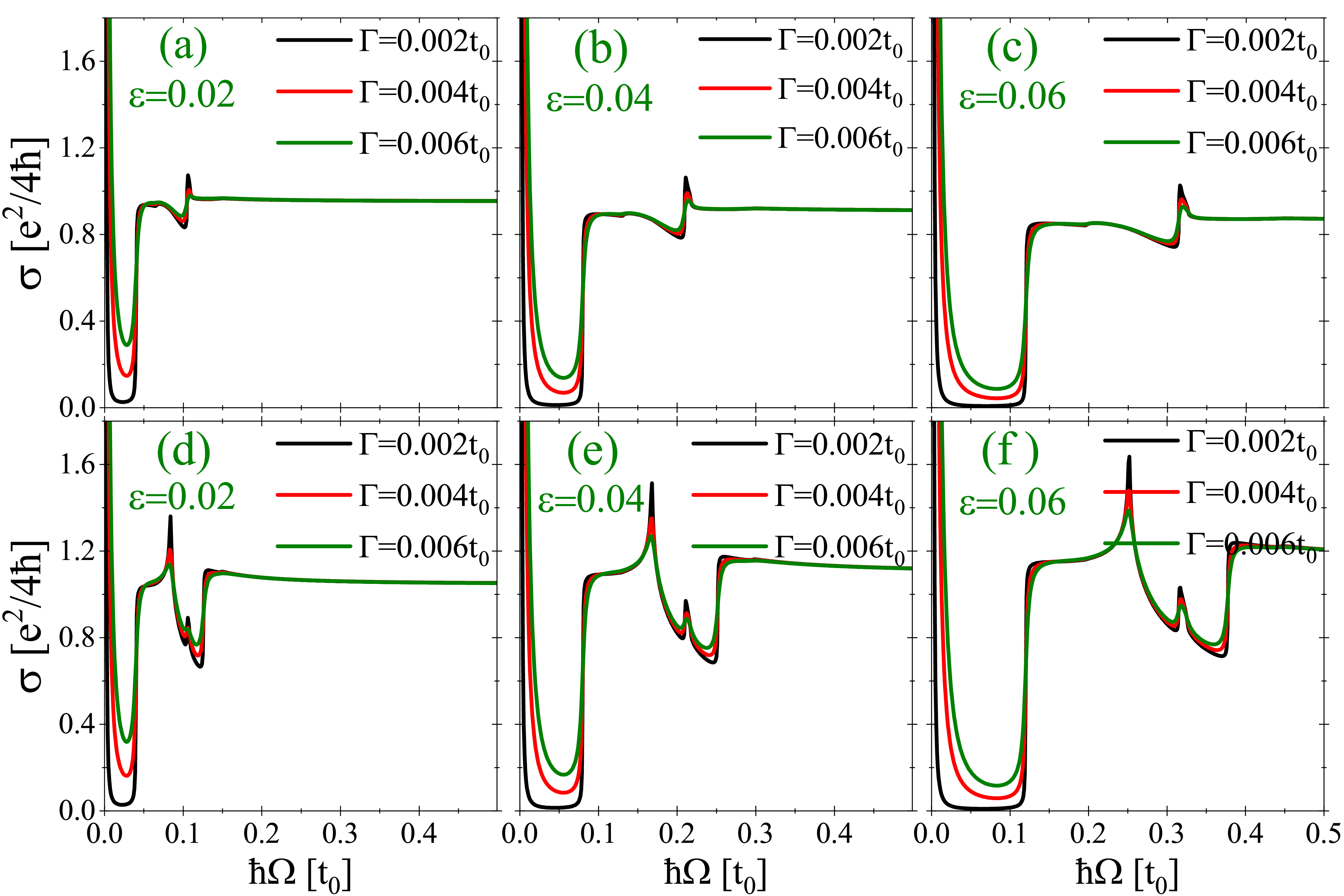}
\caption{\label{Fig08} Optical conductivity of strained graphene in the strained Kekul\'e-Y graphene, calculated for three values of the phenomenological broadening parameter, $\Gamma = 0.002~t_{0}$, $0.004~t_{0}$, and $0.006~t_{0}$, as in the main text. Other parameters are identical to those in Fig.~\ref{Fig06}. The upper and lower panels correspond to the optical response along the $x$ and $y$ directions, respectively.}
\end{figure}

To examine disorder effects, we introduce a finite lifetime broadening $\Gamma$, approximating the Dirac delta with a Lorentzian of width $\Gamma = 0.002~t_{0}$, $0.004~t_{0}$, and $0.006~t_{0}$, keeping other parameters fixed. $\sigma_{xx}$ and $\sigma_{yy}$ are displayed in upper and lower panels of Fig.~\ref{Fig08}. The inclusion of impurity scattering slightly smooths the otherwise abrupt increase in optical conductivity at $\hbar\Omega=2\mu$~\cite{Stauber3,Nair1}. Increasing $\Gamma$ smooths the spectral features and moderately reduces their intensity, but they remain clearly observable at large strain. This confirms that these features are intrinsic to strain-induced band reconstruction rather than quasiparticle lifetime effects. They serve as distinct optical fingerprints of the Kekulé-Y phase under strain.

\begin{figure}[b]
\includegraphics[width=8cm,angle=0]{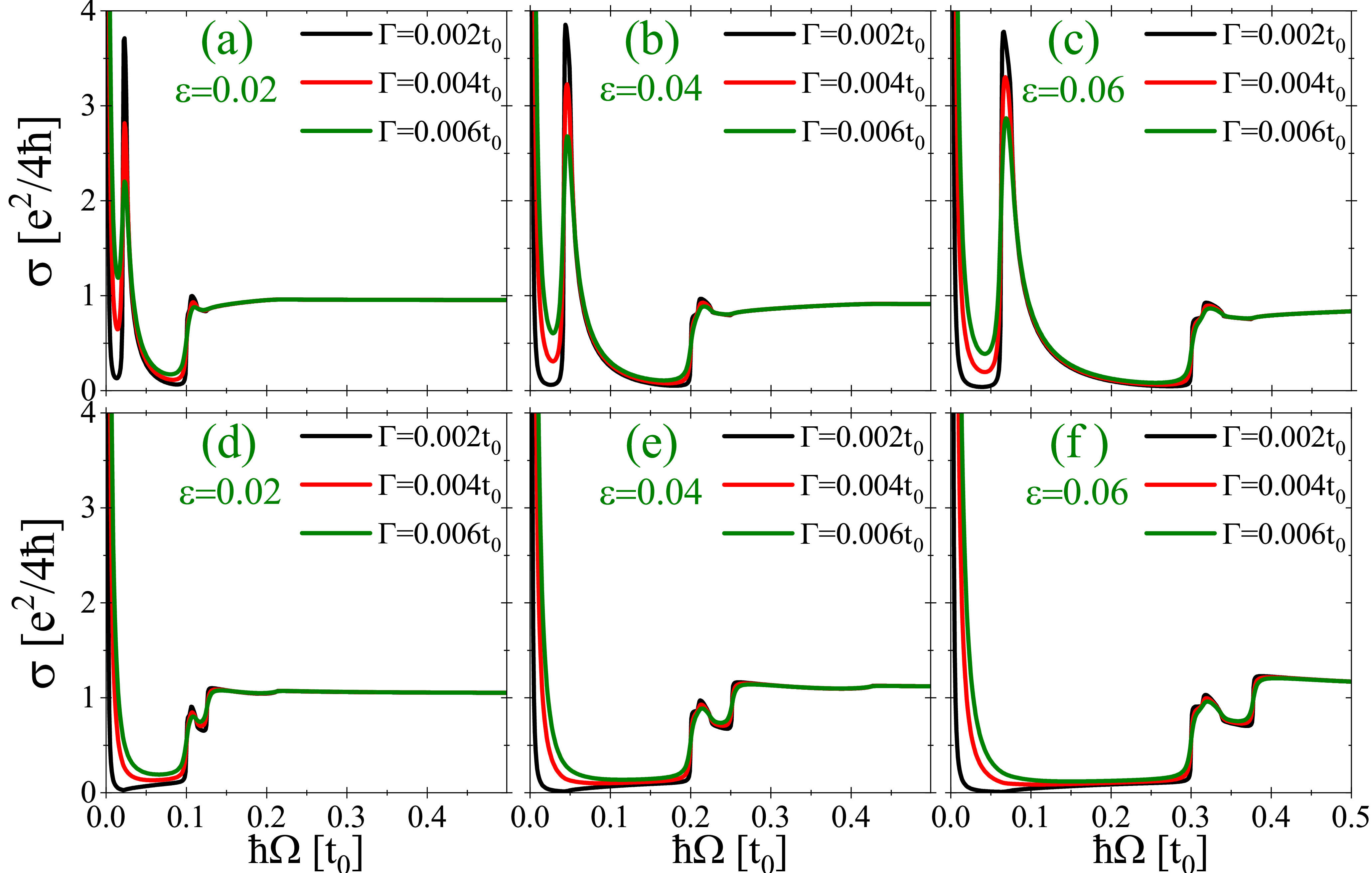}
\caption{\label{Fig09} Optical conductivity of strained Kekul\'e-Y graphene, calculated for three different chemical potentials located at the midpoint of the band gap above the saddle-point energy. The phenomenological broadening parameter takes the values $\Gamma = 0.002~t_{0}$, $0.004~t_{0}$, and $0.006~t_{0}$, while all other parameters remain the same as in Fig.~\ref{Fig06}. The upper and lower panels display the optical response along the $x$ and $y$ directions, respectively.}
\end{figure}

Fig.~\ref{Fig09}, examines the same scenario with varying chemical potential at the band-gap midpoint above the saddle point, while all other parameters are kept the same as in Fig.~\ref{Fig08}, providing a clear view of the strain-induced resonance under conditions where disorder-induced scattering effects can be distinguished. The strain-induced optical resonance position remains unchanged with increasing $\Gamma$, while intensity shows moderate reduction. Suppression becomes negligible at higher strain, demonstrating intrinsic robustness against disorder.

\begin{figure}[b]
\includegraphics[width=8cm,angle=0]{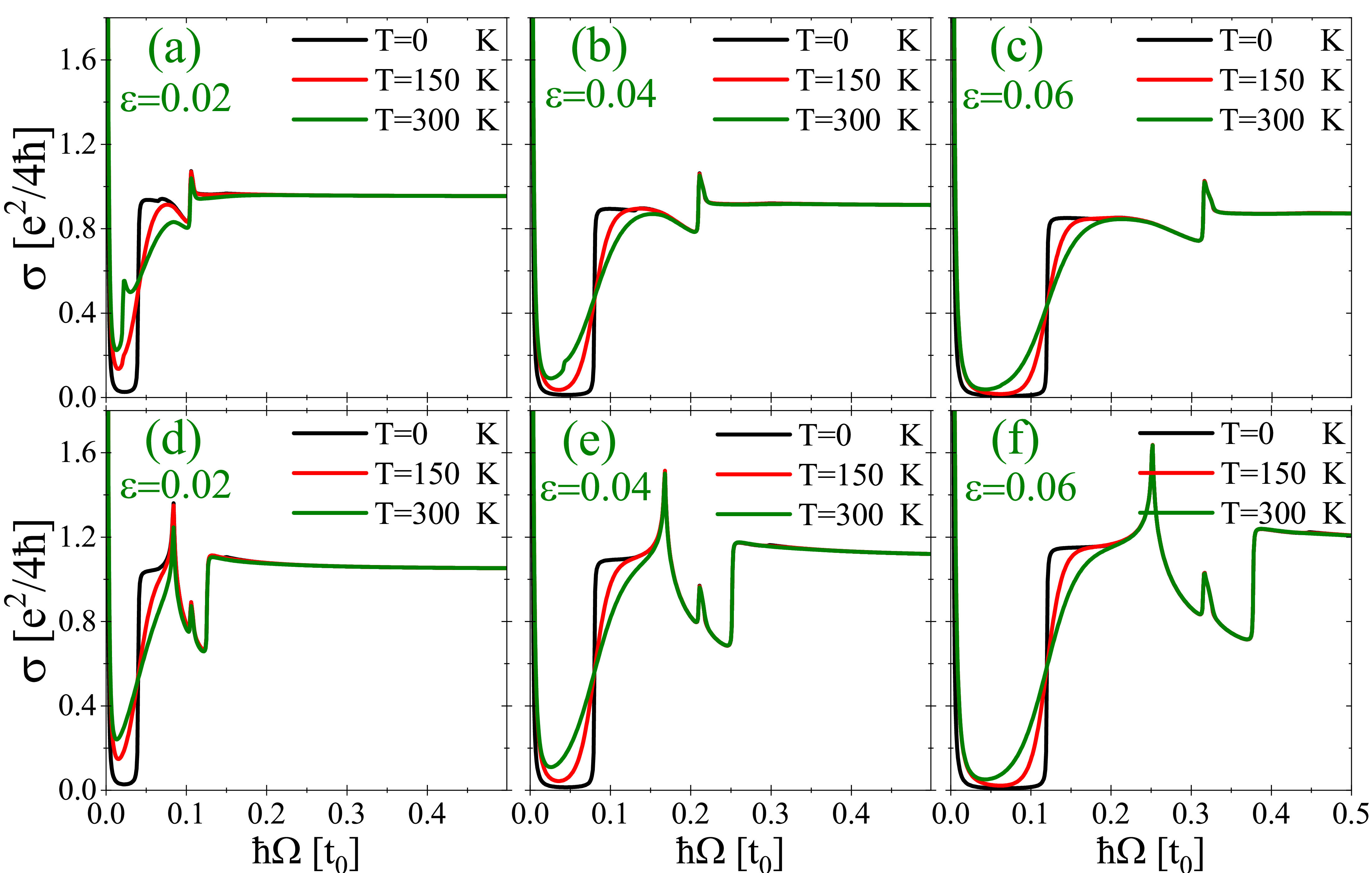}
\caption{\label{Fig10} Optical conductivity of strained Kekul\'e-Y graphene at three different temperatures, $T = 0$, 150, and 300~K, with the broadening parameter fixed at $\Gamma = 0.0005\,t_0$, but with the chemical potential set at half the saddle-point energy. The upper and lower panels show the optical conductivity along the $x$ and $y$ directions, respectively.}
\end{figure}

We now examine temperature effects. Fig.~\ref{Fig10} presents the optical conductivity at $T = 0$, 150, and 300~K, using the same parameters as in Fig.~\ref{Fig06}, with fixed broadening. $\sigma_{xx}$ is shown in the upper panels, and $\sigma_{yy}$ in the lower panels. Increasing temperature has two main effects on the optical spectra. First, the characteristic dip–peak responses are partially suppressed and broadened due to thermal broadening of the Fermi–Dirac occupation factor. Second, the sharp onset of the optical conductivity at $2\mu$, which is abrupt at $T=0$, becomes gradually broadened due to thermal smearing of the Fermi–Dirac distribution~\cite{Stauber3,Peres3}. Despite this, spectral features remain clearly identifiable even at 300 K. They are particularly robust at higher strain, confirming their value as optical fingerprints of strained Kekulé-Y graphene.

\begin{figure}[b]
\includegraphics[width=8cm,angle=0]{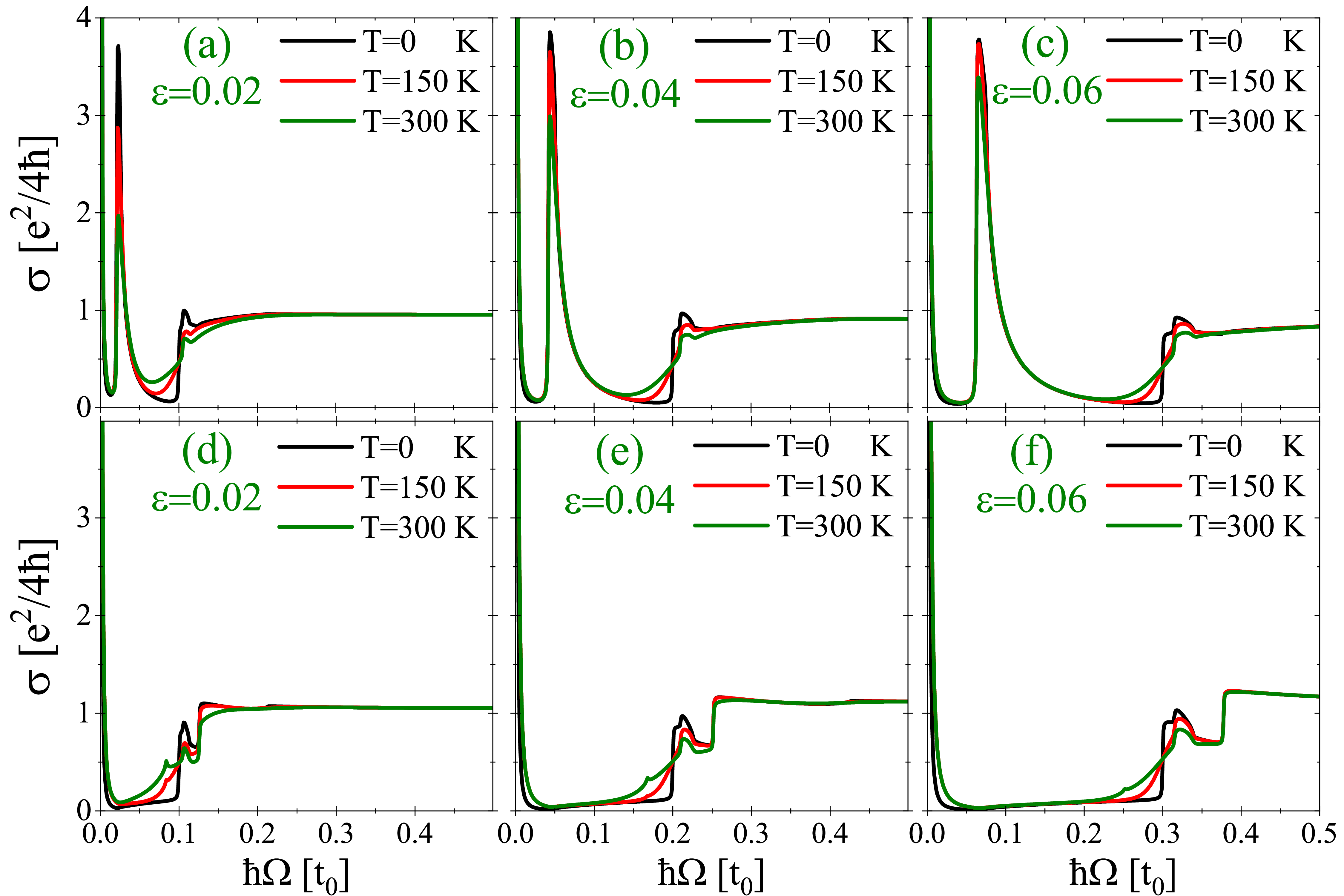}
\caption{\label{Fig11} Optical conductivity of strained Kekul\'e-Y graphene at three different temperatures, $T = 0$, 150, and 300~K, with the broadening parameter fixed at $\Gamma = 0.0005\,t_0$. Here the chemical potential is placed at the midpoint of the band gap above the saddle-point energy. The upper and lower panels show the optical conductivity along the $x$ and $y$ directions, respectively.}
\end{figure}

To further elucidate the role of the chemical potential, we now consider the optical conductivity obtained from the same data set as in Fig.~\ref{Fig10}, but with the chemical potential positioned at the midpoint of the band gap above the saddle-point energy, as shown in Fig.~\ref{Fig11}. In this configuration, some of the dip–peak structures are suppressed due to Pauli blocking, and thermal broadening further reduces their intensity and increases their width. Peak positions remain essentially unchanged, and their intensity is only weakly affected at larger strains. This confirms that the strain-induced optical resonance is highly robust against temperature, indicating its role as a robust optical signature of Kekulé-Y graphene.

Our numerical analysis demonstrates that Kekul\'e-Y graphene under uniaxial strain hosts a pronounced optical resonance peak. This peak is robust against temperature, doping, and disorder. It originates from strain-induced reconstruction of Dirac cones and anisotropic Van Hove singularities, producing sharp, direction-dependent features in the optical conductivity. The persistence and tunability of this resonance under realistic conditions make it a direct optical fingerprint of the Kekul\'e-Y phase and suggest that it should be experimentally observable using optical spectroscopy techniques.

Finally, although the experimental implementation of uniaxial strain in Kekul\'e-Y graphene has not yet been fully realized, it is important to note that Kekul\'e-Y graphene has already been successfully synthesized on Cu substrates~\cite{Gutierrez1}. Moreover, similar to other two-dimensional materials, its chemical potential can be tuned via doping~\cite{Novoselov1}. In addition to substrate-induced strain, controlled strain in graphene and other two-dimensional materials can be applied using established techniques that do not require deformation of the substrate. For example, strain can be generated by depositing stressed thin films or metallic pads on top of the graphene layer, which transfer strain through intrinsic film stress or thermal contraction~\cite{Shioya1,Shioya2}. Localized strain fields can also be induced using AFM or nano-indentation, as demonstrated in monolayer graphene~\cite{Mohiuddin1,Lee1}. These strain-engineering approaches have been widely employed to explore strain-dependent electronic and optical properties in graphene and other two-dimensional materials.

Taken together, these theoretical and experimental developments indicate that the predicted optical resonance and other characteristic features of Kekul\'e-Y graphene are likely accessible using standard optical spectroscopy techniques in the near future~\cite{Naumis1,ahn1,mennel1}. This provides promising avenues for the experimental identification of Kekul\'e-Y phases and for manipulating valley-dependent optical responses in two-dimensional materials.

\section{\label{sec:Conclusions}Conclusions}

In summary, this study presents a comprehensive analysis of the optical conductivity of Kekulé-Y graphene under uniaxial strain, investigating its spectral response at various chemical potentials, temperatures, and impurity broadenings. Strain is shown to reshape the low-energy electronic structure of the Kekulé-Y phase, inducing Van Hove singularities at energies significantly lower than those of pristine graphene. Using the Kubo formalism, we calculate the optical conductivity and identify several anisotropic low-energy interband features, with a pronounced resonance arising from strain-induced Van Hove singularities. This resonance is strongly anisotropic and remains robust against moderate thermal broadening and disorder, serving as a clear optical signature of Kekulé-Y ordering. The robustness of this resonance suggests its potential as a reliable probe for detecting strain-induced changes in the Kekulé-Y phase.

Additionally, we provide analytical expressions for the low-energy optical conductivity and the Drude weight, offering quantitative insight into the underlying band structure. These expressions emphasize how strain and thermal broadening influence the system's optical properties, providing a solid theoretical framework for interpreting experimental data.

Finally, although the controlled experimental realization of uniaxial strain in Kekul\'e-Y graphene remains challenging, recent progress in the synthesis and manipulation of two-dimensional materials suggests that such conditions are within experimental reach. The predicted anisotropic optical resonances and low-energy spectral features should therefore be accessible using standard optical spectroscopy techniques. Our results thus provide clear theoretical guidance for future experiments aimed at identifying Kekul\'e-Y ordering and exploring strain-controlled valley-dependent optical responses in graphene-based systems.

\begin{acknowledgments}
This work was supported by Farhangian University.
\end{acknowledgments}

\appendix

\section{\label{AppendixA} Construction of the low-energy Hamiltonian from the four-band spectrum}

In this Appendix, we derive the effective low-energy Hamiltonians obtained from the four-band electronic structure of Kekulé-Y graphene subjected to uniaxial strain. We show that, near the strain-induced Dirac points, the spectrum can be consistently reduced to two independent two-band models.

Under uniaxial strain, the four-band spectrum separates in the low-energy sector into two distinct groups (see Eq.~\ref{Eq06} and Fig.~\ref{Fig02}). The high-energy bands are shifted away from the Fermi level, while the low-energy bands form one two-band sector around each strain-induced Dirac point with approximately linear dispersion. This separation justifies the construction of effective low-energy Hamiltonians describing the electronic properties in the vicinity of the newly generated Dirac points.

We expand the spectrum around the strain-induced Dirac points located at $\pm \bm{A}$, under the condition that $|\bar{\bm{k}}\mp\bm{A}| \ll |\bm{A}|$. Within this parametrization, the energy bands $E^{\lambda,-}_{\bm{k}}$ around the strain-induced Dirac points can be approximated as
\begin{eqnarray}
E^{\lambda,-}_{\bm{k}} &\approx& 2 \lambda |\bm{A}| \hbar v_F \frac{\sqrt{1+\Delta^2}}{\sqrt{2}}
\Bigg[ 1 + \frac{|\bar{\bm{k}}\mp\bm{A}|^2}{4|\bm{A}|^2} \nonumber \\
&-& \Bigg( \Big[ 1 - \frac{|\bar{\bm{k}}\mp\bm{A}|^2}{4|\bm{A}|^2} \Big]^2
+ \Big[\frac{4 \Delta}{1+\Delta^2} \frac{|\bar{\bm{k}}\mp\bm{A}|}{2|\bm{A}|} \Big]^2
\Bigg)^{1/2} \Bigg]^{1/2}.
\label{EqA1}
\end{eqnarray}
Since $|\bar{\bm{k}}\mp\bm{A}| \ll |\bm{A}|$, we expand this expression in powers of $|\bar{\bm{k}}\mp\bm{A}|/|\bm{A}|$ and retain terms up to second order. Simplifying the result yields the linear dispersion relation
\begin{equation}
E^{\lambda,-}_{\bm{k}} \approx \lambda \frac{1-\Delta^2}{\sqrt{1+\Delta^2}}\, \hbar v_F |\bar{\bm{k}}\mp\bm{A}|.
\label{EqA2}
\end{equation}

The linearized dispersion, together with the symmetries preserved around each Dirac point, leads to the
following effective Hamiltonian in the vicinity of $\pm \bm{A}$:
\begin{equation}
\mathbf{H}_{\mathrm{eff}}^{\pm}(\bm{k})
= \hbar \Big(
[v_x k_x \mp v_F A_x']\, \sigma_x
+ [v_y k_y \mp v_F A_y']\, \sigma_y
\Big).
\label{EqA3}
\end{equation}
Here, the upper (lower) sign corresponds to the Dirac point at $+\bm{A}$ ($-\bm{A}$), $\sigma_{x,y}$ are Pauli matrices acting in the sublattice pseudospin space, and
$A_{x(y)}' = \frac{1-\Delta^2}{\sqrt{1+\Delta^2}}\, A_{x(y)}$. This Hamiltonian reproduces the linearized dispersion obtained above and captures the anisotropic Dirac character induced by strain.
For a uniaxial strain of magnitude $\epsilon$ applied along the zigzag direction, the renormalized Fermi velocities along the $x$ and $y$ directions are given by
\begin{subequations}
\begin{eqnarray}
v_x &= \frac{1-\Delta^2}{\sqrt{1+\Delta^2}}
\big[ 1 + (1-\beta)\varepsilon \big] v_F, \\
v_y &= \frac{1-\Delta^2}{\sqrt{1+\Delta^2}}
\big[ 1 - (1-\beta)\nu \varepsilon \big] v_F.
\end{eqnarray}
\label{EqA4}
\end{subequations}

Having established the effective low-energy description, we proceed to evaluate the interband and intraband contributions to the longitudinal optical conductivity using Eqs.~(\ref{Eq07})-(\ref{Eq11}). The effective spectral function is defined as
\begin{equation}
\bm{A}_{\mathrm{eff}}(\omega,\bm{k})
= 2\pi \sum_{\lambda}
\bm{M}_{\mathrm{eff}}^{\lambda,-}(\bm{k})\,
\delta\!\left(\hbar\omega - E_{\bm{k}}^{\lambda,-}\right).
\label{EqA5}
\end{equation}
The corresponding band-projection matrix reads
\begin{equation}
\bm{M}_{\mathrm{eff}}^{\lambda,-}(\bm{k})=
\begin{pmatrix}
M_{11} & \lambda M_{12} \\
\lambda M_{12}^{\ast} & M_{11}
\end{pmatrix},
\label{EqA6}
\end{equation}
where $M_{11}=\frac{1}{2}$, and
\begin{equation}
M_{12}=\frac{\hbar \big( \big[ v_{x} k_{x}\mp v_F A_{x}^{'} \big] + i \big[ v_{y} k_{y}\mp v_F A_{y}^{'} \big] \big)}{2E_{\mathbf{k}}^{+,-}}.
\label{EqA7}
\end{equation}

Substituting these expressions into Eq.~(\ref{EqB8}), we obtain the interband squared transition matrix element,
\begin{equation}
P_{\mathrm{eff},\alpha\alpha}^{-,-\rightarrow +,-}(\bm{k}) =\frac{1}{2} \Big(\frac{v_{\alpha} \big[ v_{\bar{\alpha}} k_{\bar{\alpha}}\mp v_F A_{\bar{\alpha}}^{'} \big]}{E_{\bm{k}}^{+,-}}  \Big)^{2},
\label{EqA8}
\end{equation}
where $\bar{\alpha}$ is the direction perpendicular to $\alpha$ ($x \leftrightarrow y$).

By performing a change of variables where \( v_{\alpha} k_{\alpha} \mp v_F A^{'}_{\alpha} \) is replaced by \( q_{\alpha} \), and substituting Eq.~(\ref{EqA8}) into the expressions in Eq.~\ref{Eq11}, we integrate over the \( q_x \) and \( q_y \) coordinates to obtain the optical conductivity, i.e., the relation given by Eq.~\ref{Eq17}. Furthermore, by substituting these expressions into Eq.~\ref{Eq13}, we derive the corresponding Drude weight relations, i.e., those given by Eq.~\ref{Eq14} and Eq.~\ref{Eq16}.

\section{\label{AppendixB} Derivation of the optical conductivity within linear-response theory}

In this Appendix, we present the intermediate steps leading to the expressions for the optical conductivity used in the main text. The derivation is carried out within linear-response theory and follows a standard Kubo-formula approach. To this end, we evaluate the current–current correlation function \( \Pi_{\alpha\alpha}(\Omega + i0^{+}) \).

Within linear-response theory, the function \( \Pi_{\alpha\alpha}(\Omega + i0^{+}) \) is obtained by performing the analytical continuation of the corresponding Matsubara current–current correlation function \( (i\Omega_n \rightarrow \Omega + i0^{+}) \), defined as~\cite{Stauber1}
\begin{equation}\label{EqB1}
\Pi_{\alpha\alpha}(i\Omega_n) = \int_0^{\hbar\beta} d\tau e^{i\Omega_n \tau} \left\langle T_{\tau} j_{\alpha}^{P}(\tau) j_{\alpha}^{P}(0) \right\rangle_0,
\end{equation}
where \( i\Omega_n \) denote the bosonic Matsubara frequencies, \( \beta = \frac{1}{k_B T} \), and \( T_\tau \) is the imaginary-time ordering operator. The paramagnetic current operator is given by~\cite{Nicol1}
\begin{equation}\label{EqB2}
j^{P}_{\alpha} = - e \sum_{\mathbf{k}} \bm{\psi}^{\dagger}_{\mathbf{k}} \, \bm{v}_{\alpha} \, \bm{\psi}_{\mathbf{k}},
\end{equation}
where \( e \) is the electron charge, and the velocity operator is defined as \( \bm{v}_{\alpha} = \frac{\partial \bm{H}}{\hbar \, \partial k_{\alpha}} \). Substituting this operator into the definition of the correlation function and applying Wick’s theorem, the correlation function within the noninteracting (bubble) approximation can be written as~\cite{Nicol1}
\begin{eqnarray}\label{EqB3}
\Pi_{\alpha\alpha}(i\Omega_n) &=& \frac{g_s \hbar e^2}{\beta} \sum_{i\nu_m} \sum_{\mathbf{k}} \nonumber \\
 &\textrm{Tr}& \left[ \bm{v}_{\alpha} \bm{G}(i\nu_m + i\Omega_n, \mathbf{k}) \bm{v}_{\alpha} \bm{G}(i\nu_m, \mathbf{k}) \right],
\end{eqnarray}
where \( g_s = 2 \) is the spin degeneracy, \( i\nu_m \) are the fermionic Matsubara frequencies, and \( \bm{G}(i\nu_m, \mathbf{k}) = \left( i\hbar\nu_m \bm{1} - \bm{H} \right)^{-1} \) is the single-particle Green's function. The trace is taken over the internal band (or pseudospin) degrees of freedom. To proceed, we express the Green’s function in its spectral representation,
\begin{equation}\label{EqB4}
\bm{G}(z, \mathbf{k}) = \int_{-\infty}^{+\infty} \frac{d\omega}{2\pi} \frac{\bm{A}(\omega, \mathbf{k})}{z - \omega}.
\end{equation}
Substituting this representation into the correlation function and performing the summation over fermionic Matsubara frequencies, one obtains
\begin{eqnarray}\label{EqB5}
\Pi_{\alpha\alpha}(i\Omega_n) &=& g_s \hbar^2 e^2 \int_{-\infty}^{+\infty} \frac{d\omega}{2\pi} \int_{-\infty}^{+\infty} \frac{d\omega'}{2\pi} \sum_{\mathbf{k}} \frac{f_{\omega} - f_{\omega'}}{i\Omega_n + \omega - \omega'} \nonumber \\
 &\textrm{Tr}& \left[ \bm{v}_{\alpha} \bm{A}(\omega', \mathbf{k}) \bm{v}_{\alpha} \bm{A}(\omega, \mathbf{k}) \right],
\end{eqnarray}
where \( f_{\omega} = \left[ e^{\beta (\hbar\omega - \mu)} + 1 \right]^{-1} \) is the Fermi-Dirac distribution function, \( \mu \) is the chemical potential. The spectral matrix can be expressed as
\begin{equation}\label{EqB6}
\bm{A}(\omega, \mathbf{k}) = 2\pi \sum_{\lambda, s} \bm{M}^{\lambda, s}(\bm{k})~ \delta(\hbar \omega - E_{\mathbf{k}}^{\lambda, s}),
\end{equation}
where \( \ \bm{M}^{\lambda, s}(\bm{k}) \) denotes the projection operator onto the eigenstate labeled by band index \( (\lambda,s) \).

Upon substituting this expression, carrying out the frequency integrations, and performing the analytical continuation, \( i\Omega_n \rightarrow \Omega + i0^{+} \), the current–current correlation function takes the form
\begin{widetext}
\begin{equation}\label{EqB7}
\Pi_{\alpha\alpha}(\Omega + i0^{+})=-g_{s}\hbar e^{2}\sum_{\mathbf{k}}\sum_{\lambda,\lambda^{'},s,s^{'}}
\frac{f_{\mathbf{k}}^{\lambda,s} - f_{\mathbf{k}}^{\lambda',s'}}
{E_{\mathbf{k}}^{\lambda,s}-E_{\mathbf{k}}^{\lambda^{'},s^{'}}+\hbar\Omega+i0^{+}}
P_{\alpha\alpha}^{\lambda,s \to \lambda',s'}(\mathbf{k}),
\end{equation}
\end{widetext}
where the squared transition matrix element of the current operator is defined as
\begin{equation}\label{EqB8}
P_{\alpha\alpha}^{\lambda,s \to \lambda',s'}(\mathbf{k}) = \textrm{Tr} \left[ \bm{v}_{\alpha}~\bm{M}^{\lambda', s'}(\bm{k})~\bm{v}_{\alpha}~\bm{M}^{\lambda, s}(\bm{k}) \right].
\end{equation}

\nocite{*}

\bibliography{Refs}% Produces the bibliography via BibTeX.
%\bibliography{Manuscript.bbl}

\end{document}